\pdfoutput=1
\documentclass{article}

\usepackage{iclr2027_conference,times}
\iclrfinalcopy

\usepackage[utf8]{inputenc}
\usepackage[T1]{fontenc}

\usepackage{microtype}
\usepackage[hidelinks]{hyperref}
\usepackage{url}
\usepackage{booktabs}
\usepackage{amsfonts}
\usepackage{nicefrac}
\usepackage{xcolor}
\usepackage{booktabs}
\usepackage{algorithm}
\usepackage{algpseudocode}

\usepackage{amsmath,amssymb}
\usepackage{graphicx}
\usepackage{enumitem}
\usepackage{placeins}
\usepackage{float}

 \renewcommand{\tilde}{\widetilde}
 \renewcommand{\hat}{\widehat}

 \newcommand{\tvec}[1]{\ensuremath{\Tilde{\boldsymbol{#1}}}}

 \renewcommand{\vec}[1]{\ensuremath{\boldsymbol{#1}}}

 \newcommand{\Real}{{\mathbb{R}}}

 \DeclareMathOperator{\diag}{diag}

 \renewcommand{\eqref}[1]{(\ref{eq:#1})}

 \newcommand{\figref}[1]{Fig.~\ref{fig:#1}}
 \newcommand{\tabref}[1]{Table~\ref{tab:#1}}
 \newcommand{\secref}[1]{Section~\ref{sec:#1}}
 \newcommand{\appref}[1]{Appendix~\ref{app:#1}}

 \newcommand{\PoDAR}{_{\text{PoDAR}}}
 
\title{PoDAR: Power-Decoupled Audio\\Representation for Generative Modeling}

\author{%
  Alejandro Luebs \\
  Descript \\
  \texttt{alejandroluebs@gmail.com} \\
  \And
  Mithilesh Vaidya \\
  Descript \\
  \texttt{mithilesh@descript.com} \\
  \And
  Ishaan Kumar \\
  Descript \\
  \texttt{ishaan@descript.com} \\
  \And
  Sumukh Badam \\
  Descript \\
  \texttt{sumukh@descript.com} \\
  \AND
  Stephen W. Bailey \\
  Descript \\
  \texttt{stephen@descript.com} \\
  \And
  Matthew Bendel \\
  Descript \\
  \texttt{matt.bendel@descript.com} \\
  \And
  Jose Sotelo \\
  Descript \\
  \texttt{rdz.sotelo@gmail.com} \\
  \And
  Xingzhe He \\
  Descript \\
  \texttt{xingzhe@descript.com} \\
}

\begin{document}

\maketitle
\lhead{Preprint}

\begin{abstract}
The performance of audio latent diffusion models is primarily governed by generator expressivity and the modelability of the underlying latent space. While recent research has focused primarily on the former, as well as improving the reconstruction fidelity of audio codecs, we demonstrate that latent modelability can be significantly improved through explicit factor decoupling.
We present PoDAR (\textbf{Po}wer-\textbf{D}ecoupled \textbf{A}udio \textbf{R}epresentation), a framework that utilizes a randomized power augmentation and latent consistency objective to decouple signal power from invariant semantic content.
This factorization makes the latent space easier to model, which both accelerates the convergence of downstream generative models and improves final overall performance.
When applied to a Stable Audio 1.0 VAE with an F5-TTS generator, PoDAR achieves about a $2\times$ acceleration in convergence to match baseline performance, while increasing final speaker similarity by 0.055 and UTMOS by 0.22 on the LibriSpeech-PC dataset.
Furthermore, isolating power into dedicated channels enables the application of CFG exclusively to power-invariant content, effectively extending the stable guidance regime to higher scales.

\end{abstract}

\begin{figure}[htbp]
    \centering
    % First plot
    \includegraphics[width=0.9\textwidth]{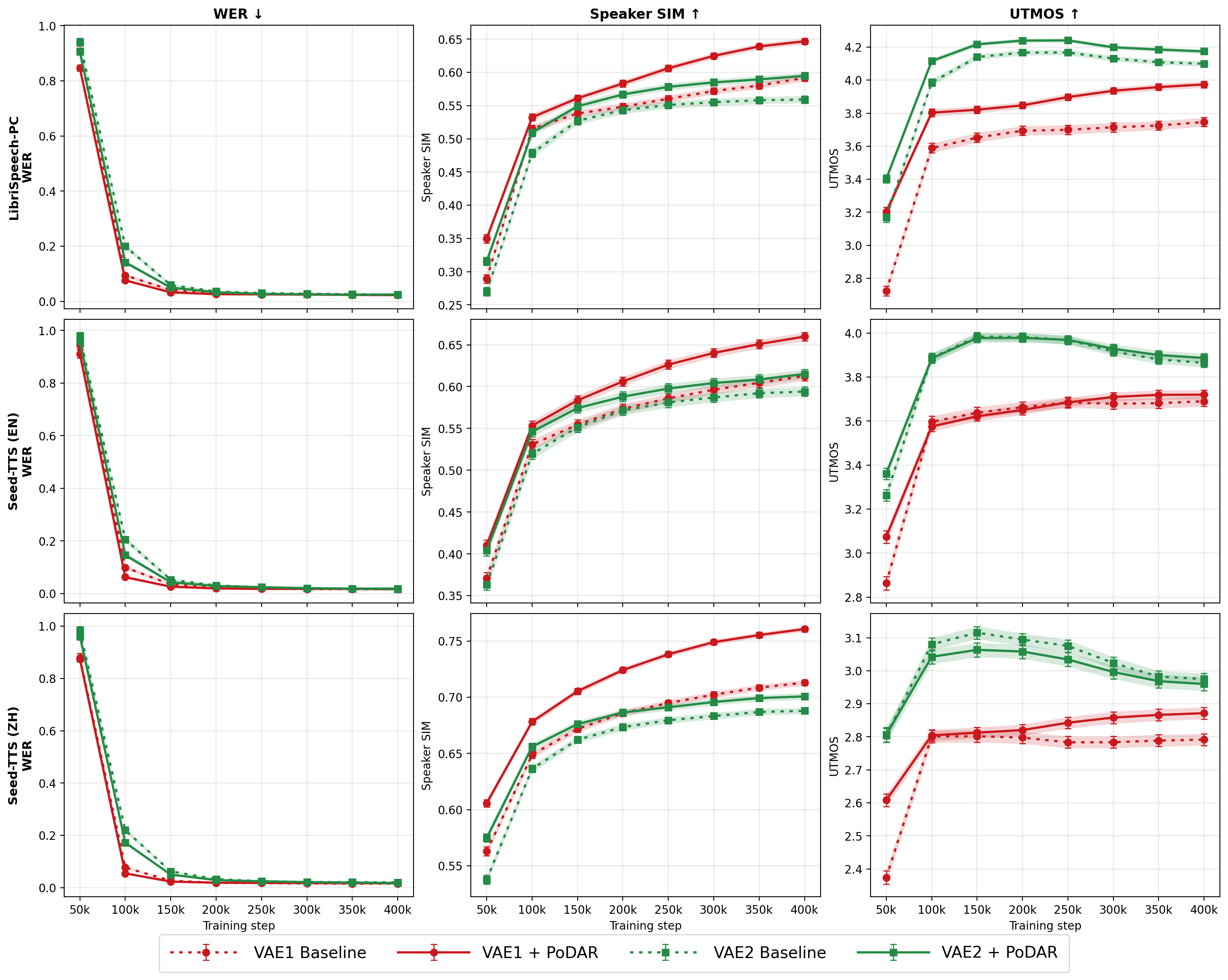}
    \caption{Comparison of WER, speaker SIM, and UTMOS for baseline and PoDAR configurations as a function of training progress for the LibriSpeech and Seed-TTS datasets. We observe that generators trained on codecs with PoDAR consistently outperform their baseline counterparts across speaker similarity (SIM) and speech quality (UTMOS). VAE1 and VAE2 refer to standard codecs from \texttt{stable-audio-tools}~\citep{stabilityai_stableaudiotools}, based on DAC~\citep{kumar2023rvqgan} and Oobleck respectively.}
    \label{fig:all}
\end{figure}

\section{Introduction} \label{sec:introduction}

Modern high-fidelity image and audio generators often adopt a two-stage latent modeling pipeline where an autoencoder maps high-dimensional signals to a compact continuous representation and a downstream diffusion or flow model learns the distribution of these latents under text or other conditioning. Besides avoiding the cost of modeling raw signals, latent representations provide smoother and more semantically structured manifolds that are easier to model than the pixel or waveform space \citep{rombach2022ldm,liu2023audioldm,evans2024stableaudioopen}. Within this framework, the autoencoder functions as more than a simple compressor because it establishes the fundamental geometry and statistics that the downstream generator must ultimately navigate.

This design makes the \emph{modelability} of the latent space a central concern, where a space is considered modelable if a downstream generator can learn its distribution efficiently while maintaining fast convergence and high quality under guided conditioning~\citep{rombach2022ldm}. Early autoencoders utilized in latent generative modeling were optimized primarily for reconstruction fidelity and perceptual quality, or compression ratio~\citep{rombach2022ldm, esser2021taming}. However, a representation that achieves high reconstruction accuracy is not necessarily easy to model~\citep{xu2026makingreconstructionfidpredictive}.

Methods such as \citet{leng2025repa, zheng2025diffusion, defossez2024moshi, li2025dualcodec} report improved performance of downstream generator modelability by using a frozen pre-trained encoder to infuse semantic information into the latent space. However, the primary drawback of these methods is that selecting an appropriate representation is highly non-trivial, often requiring extensive ablations \citep{yu2024representation} over encoder architectures and intermediate feature layers. Encoder quality does not necessarily correlate with downstream generative performance \citep{singh2025matters}, evidenced by the fact that alignment with larger variants within the same model family can result in similarly or worse generative performance \citep{didi2025r4g}. Additionally, these approaches increase the memory footprint during training.

On the other hand, our experiments suggest that semantic information becomes entangled with independent nuisance variables, such as signal power, which are largely irrelevant to the linguistic content of a speech signal. When such attributes are entangled with semantic content in the latent space, the downstream generator must learn a complex joint distribution over unrelated factors, thereby impairing latent modelability.

Considering this, we take an orthogonal approach to improving the modelability of the latent space, which can be deployed either as a standalone alternative or in conjunction with representation alignment. Power-Decoupled Audio Representation (PoDAR) is a lightweight, representation-agnostic, self-supervised framework that partitions the autoencoder latent space into a power subspace and a power-invariant semantic subspace, without relying on external pre-trained encoders. During the autoencoder training, we apply randomized power perturbations and employ a consistency loss to enforce this decoupling. The primary goal of this framework is not to obtain interpretable factors, but rather to apply these techniques to directly address the modelability bottleneck in two-stage generation by producing a latent distribution that is significantly easier for a downstream diffusion or flow model to learn. Our empirical results show that this representation-level structure improves the final generation quality and speeds up convergence by around $2\times$, as illustrated in \figref{all}.

The same factorization also provides a natural modification for Classifier-Free Guidance~\citep[CFG;][]{ho2022cfg}. Within a conventional entangled space, any attempt to boost conditional adherence may inadvertently amplify nuisance variables, such as signal power, and distort the signal in the process. In a power-decoupled latent space, however, guidance can be applied selectively to the semantic subspace, while leaving the power channels governed strictly by the learned generative dynamics. Improved robustness to high guidance scales thus emerges as an advantageous secondary effect of this method.

\paragraph{Our contributions.}
The following points summarize the primary technical and empirical contributions of this work:
\begin{enumerate}
    \item \textbf{The PoDAR framework} establishes a lightweight, architecture-agnostic, self-supervised approach for learning power-decoupled audio representations by utilizing a randomized power augmentation and latent consistency objective to partition the latent space into power and power-invariant semantic subspaces without relying on external pre-trained encoders.
    \item \textbf{Enhanced latent modelability} results from the explicit decoupling of signal power from semantic content, allowing generators to converge about $2\times$ faster compared to conventional entangled baselines, while achieving superior final results.
    \item \textbf{Partial CFG} allows steering exclusively over the semantic subspace, improving robustness at higher guidance scales.
\end{enumerate}

\section{Background} \label{sec:background}

\paragraph{Latent generative modeling.}
Modern high-fidelity generative modeling often utilizes a learned continuous latent representation rather than operating directly on raw pixels or waveforms~\citep{rombach2022ldm,liu2023audioldm,evans2024stableaudioopen}. Let $\vec{x} \in \mathbb{R}^N$ represent an input signal, such as an image tensor or audio waveform. An autoencoder establishes an encoder-decoder pair through the following transformation:
\begin{equation}
z = E(x),
\qquad
\hat{x} = D(z)
\end{equation}
where $L$ denotes the number of latent dimensions, $T$ indexes spatial or temporal locations and $z\in \mathbb{R}^{L \times T}$ denotes the latent. Following this initial stage, a downstream conditional generator is trained to learn the distribution $p(\vec{z} \mid \text{cond})$ where $\text{cond}$ signifies conditioning information, such as text or class labels. During the generation process, samples are produced in the latent space and subsequently mapped back to the signal domain using the pre-trained decoder $\vec{D}(\cdot)$.

Diffusion and flow-based models provide robust mechanisms for learning complex distributions $p(\vec{z} \mid \text{cond})$. Diffusion models generate samples by iteratively denoising an initial noisy variable through learned score, noise or velocity predictions \citep{ho2020ddpm,song2021scorebased,karras2022edm}, while flow matching and rectified-flow methods instead learn vector fields or ODEs that transport a base distribution to the target data distribution \citep{lipman2023flow,liu2023flow}. Regardless of the specific generative mechanism, the distribution induced by the autoencoder remains the fundamental object that the downstream generator must learn.

\paragraph{Representation alignment.}
Recent research, particularly in the image domain, has focused on improving modelability of the latent space by aligning it with a pre-trained encoder. REPA \citep{yu2024representation} improves the training of diffusion transformers by aligning internal denoising representations with pretrained visual features, while REPA-E \citep{leng2025repa} extends this concept to the end-to-end tuning of VAE tokenizers. Furthermore, Representation Autoencoders~\citep[RAEs;][]{zheng2025diffusion} replace traditional reconstruction-only VAEs with architectures built around pretrained encoders such as DINO~\citep{caron2021emerging}, SigLIP~\citep{zhai2023sigmoid} or MAE~\citep{he2022masked} to yield semantically rich latents that accelerate the convergence of diffusion models. Although these advancements were primarily developed for the image domain, there has been parallel progress on the audio front. In the development of the Moshi codec \citep{defossez2024moshi}, the authors align the primary codebook latents with a pre-trained WavLM encoder \citep{chen2022wavlm} and similarly, the DualCodec framework \citep{li2025dualcodec} aligns its representations with a pre-trained w2v-BERT-2.0 model \citep{barrault2023seamless}.

\paragraph{Factor decoupling in audio representations.}
There has been some recent work exploring decomposing neural audio representations to isolate specific acoustic attributes. Shape-gain decomposition \citep{sadok2026equalizer} has been proposed to improve codec scale-robustness and compression efficiency by manually extracting a scalar gain prior to encoding. Other approaches rely on cross-reconstruction swap losses \citep{grundhuber2026acoustic, omran2023disentangling} to separate speech from environmental features, such as reverberation or background noise, thereby enabling applications like acoustic teleportation and speech enhancement. While these methods share an interest in representation decoupling, they are fundamentally optimized for enhancement and compression tasks. In contrast, the objective of PoDAR is to decouple signal power to eliminate nuisance variables and improve the modelability of the latent space for downstream generative models.

\paragraph{Classifier-free guidance.}
Classifier-Free Guidance~\citep[CFG;][]{ho2022cfg} is a fundamental component of modern generative models, since it not only improves conditional adherence, but also generation quality \citep{karras2024guiding}. It achieves this by extrapolating from an unconditional baseline toward a conditional prediction:
\begin{equation}
\vec{v}_{\mathrm{cfg}} = \vec{v}_0 + w(\vec{v}_{\mathrm{cond}} - \vec{v}_0),
\end{equation}
where $\vec{v}_0$ is the unconditional prediction, $\vec{v}_{\mathrm{cond}}$ the conditional prediction, and the scalar $w$ controls the strength of the extrapolation towards the conditioning signal.

\section{Method}
\label{sec:method}
Our approach has two components: (i) an augmentation and consistency constraint that encourages power decoupling in latent space, and (ii) a partial CFG rule that applies guidance only to the power-invariant subspace at generation time.

\subsection{Power augmentation}
For an audio waveform $\vec{x} \in \Real^N$, we define a randomized \emph{power augmentation} $\mathcal{A}(\cdot)$
\begin{equation}
\tvec{x} = \mathcal{A}(\vec{x}).
\end{equation}
The augmentation is chosen to vary the power while preserving semantic content, applying a random global gain between $-6$ and $+6$~dB:
\begin{equation}
\tvec{x} = \mathcal{A}(\vec{x}) = g \vec{x}, \qquad g = 10^{u/20}, \qquad u \sim \mathrm{Uniform}[-6, 6].
\end{equation}
We specifically target signal power because it can be perturbed by a purely linear transformation that perfectly preserves the original waveform's temporal alignment. This exact structural correspondence is a strict prerequisite for the consistency objective defined in Section~\ref{ssec:consistency}. Attempting to isolate other acoustic features, such as pitch, voicing, or speaker ID, would typically violate the alignment requirement or introduce artifacts, which would have to be modeled by the network as well.

\subsection{Consistency constraint}
\label{ssec:consistency}
Let $\vec{z} = \vec{E}(\vec{x}) \in \Real^{L \times T}$ be the encoder output. We reserve the first $k$ channels for power and treat the remaining channels as power-invariant content:
\begin{equation}
\vec{z} = \big[\vec{z}_{p};~ \vec{z}_{c}\big], \qquad \vec{z}_{p} \in \Real^{k \times T}, \quad \vec{z}_{c} \in \Real^{(L-k)\times T}.
\end{equation}
During training, we enforce invariance of the content channels under power augmentation by penalizing deviations between content latents of $\vec{x}$ and $\tvec{x}$:
\begin{equation}
\mathcal{L}\PoDAR
= \big\| \phi(\vec{E}(\vec{x}))_{c} - \phi(\vec{E}(\tvec{x}))_{c} \big\|_2^2,
\label{eq:lcons}
\end{equation}
where $(\cdot)_{c}$ denotes channels $k, k{+}1, \dots, L{-}1$.

For deterministic autoencoders, we use $\phi(\vec{E}(\vec{x})) = \vec{E}(\vec{x})$.
For VAEs with encoder $q_\psi(\vec{z}\mid \vec{x})=\mathcal{N}(\vec{\mu}_\psi(\vec{x}), \diag(\vec{\sigma}^2_\psi(\vec{x})))$, we apply the constraint only to the mean, i.e., $\phi(\vec{E}(\vec{x}))=\vec{\mu}_\psi(\vec{x})$, justified by the latent variance analysis provided in \appref{variance}.

The full autoencoder objective becomes:
\begin{equation}
\label{eq:final}
\mathcal{L}_{\text{AE}}^{\text{new}} \;=\; \mathcal{L}_{\text{AE}}^{\text{old}} \;+\; \lambda\PoDAR\, \mathcal{L}\PoDAR.
\end{equation}
This additional loss encourages the encoder to route augmentation-induced power variation into $\vec{z}_p$ while keeping only semantic content in $\vec{z}_c$. The hyperparameter $k$ controls the capacity allocated to power, and $\lambda\PoDAR$ controls the strength of the decoupling.

\subsection{Partial CFG}
\label{ssec:partial}
We propose a modified CFG rule, which applies guidance only to the power-invariant channels:
% \begin{align}
% \vec{v}_{p} &= \vec{v}_{\text{cond},p}, \\
% \vec{v}_{c} &= \vec{v}_{0,c} + w (\vec{v}_{\text{cond},c} - \vec{v}_{0,c}),
% \end{align}
% or equivalently,
\begin{equation}
\vec{v} = \big[\vec{v}_{\text{cond},p};~ \vec{v}_{0,c} + w (\vec{v}_{\text{cond},c} - \vec{v}_{0,c})\big],
\label{eq:partialcfg}
\end{equation}
where $\vec{v}_{*,p}$ denotes channels $0, 1, \dots, k{-}1$, and $\vec{v}_{*,c}$ denotes channels $k, k{+}1, \dots, L{-}1$ of the prediction $\vec{v}_{*}$. By restricting guidance to the power-invariant subspace, partial CFG improves the robustness of the generator at elevated guidance scales.

\section{Experimental setup}
\label{sec:experiments}

\subsection{Baselines}
\label{ssec:autoencoder_baseline}
\paragraph{Autoencoder baseline.}

We evaluate two continuous audio representations derived from the
\emph{Stable Audio 1.0 VAE} (VAE1) and \emph{Stable Audio 2.0 VAE} (VAE2) autoencoders~\citep{evans2024fasttiming,evans2024stableaudioopen} implemented in
\texttt{stable-audio-tools} \citep{stabilityai_stableaudiotools}, re-training both VAEs from scratch under the original published recipe.

Both baselines are waveform autoencoders trained at 44.1~kHz with a 64-dimensional continuous Gaussian VAE bottleneck. Training is consistent across the two, with a perceptually weighted multi-resolution STFT reconstruction loss \citep{yamamoto2020parallelwavegan}, an EnCodec-style multi-scale STFT discriminator \citep{defossez2022encodec}, the same adversarial and feature-matching weights (0.1 / 5.0), and EMA.
\tabref{stable_audio_vaes} summarizes the key configuration differences between VAE1 and VAE2.

We use the same training mixture as \emph{DAC}~\citep{kumar2023rvqgan}, spanning speech: DAPS~\citep{mysore2015daps}, DNS Challenge 4 clean speech segments~\citep{dubey2022dns}, Common Voice~\citep{ardila2019commonvoice}, and VCTK~\citep{yamagishi2019vctk}; music: MUSDB18~\citep{rafii2017musdb18} and MTG-Jamendo~\citep{bogdanov2019mtgjamendo}; and environmental audio: AudioSet balanced and unbalanced train segments~\citep{gemmeke2017audioset}.

Unless otherwise noted, all representations are trained with a single power channel ($k=1$) and decoupling weights of $\lambda_{\text{PoDAR}}=0.5$ for VAE1 and $\lambda_{\text{PoDAR}}=0.1$ for VAE2, as detailed in Section~\ref{ssec:codec}.

\paragraph{Generator baseline.}

We use F5-TTS \citep{chen2025f5tts} as the text-to-speech generator,
adapted to operate on the continuous autoencoder latents introduced in
Section \ref{ssec:autoencoder_baseline}. We follow the published F5-TTS v1 Base recipe and
modify the input/output projections to match the autoencoder latent dimension. To improve
stability and training efficiency, we apply per-channel $z$-score normalization, following common latent-standardization practice in latent diffusion models \citep{rombach2022ldm}. The generator operates on normalized latents $\tilde{\vec{z}}=(\vec{z}-\vec{\mu})/(\vec{\sigma}+10^{-6})$, and we invert this transform prior to decoding.

Generators are trained on the bilingual \emph{Emilia ZH-EN} subset of the Emilia dataset \citep{he2024emilia}, as released for the original F5-TTS recipe. We use the official Emilia train split.

Unless stated otherwise, we use $w=3.0$ for VAE1-based generators and $w=2.0$ for VAE2-based
generators, which we found to be the best CFG operating point for each family
in the full CFG sweep.

More details about the exact architecture and inference strategy can be found in Table \ref{tab:hparams}.

\subsection{Metrics}
\label{ssec:metrics}
\paragraph{Reconstruction metric.} For evaluating the reconstruction quality of codecs, we report ViSQOL~\citep[audio mode;][]{hines2015visqol}, a full-reference intrusive metric based on spectro-temporal similarity, computed on the same held-out DAPS speakers (F10 and M10) used by \emph{DAC} \citep{kumar2023rvqgan}.

\paragraph{Measuring modelability.} 
To quantify the modelability of the latent space, we initially explored several direct statistical and geometric measures, including the condition number of the latent covariance matrix, the latent smoothness (described in full in \appref{latent-smoothness}), and temporal locality. Whil improvement of about 0.05 for VAE1 across e these metrics effectively captured large improvements in modelability, they lacked the sensitivity necessary to consistently reflect more modest gains. Ultimately, the definition of modelability is the difficulty of learning the underlying score or velocity fields. Therefore, we evaluate it by measuring the convergence speed and final generation metrics of the downstream flow-matching model.

\paragraph{Generation metrics.}
We evaluate generated speech following the same protocol mentioned in F5-TTS \citep{chen2025f5tts}, which includes WER and speaker similarity (SIM-o). Additionally, to assess the quality of the generated speech in a reference-free manner, we use UTMOS~\citep{saeki2022utmos}.

Results are reported on three public test sets: the LibriSpeech-PC test-clean \emph{cross-sentence} list~\citep{meister2023librispeechpc} ($1{,}127$ utterances), where LibriSpeech-PC is LibriSpeech~\citep{panayotov2015librispeech} with restored punctuation and casing; Seed-TTS test-en ($1{,}088$ utterances); and Seed-TTS test-zh~\citep{anastassiou2024seedttsfamilyhighqualityversatile} ($2{,}020$ utterances), whose Mandarin material derives from DiDiSpeech~\citep{guo2021didispeechlargescalemandarin}.

% \paragraph{Statistical significance.}
% For all plots and tables, error bars denote 95\% confidence intervals of the mean, $\bar{x} \pm 1.96\,\sigma/\sqrt{n}$, where $\sigma$ is the sample
% standard deviation and $n$ is the number of test-set utterances.

\subsection{Swap test}
\label{ssec:swap}
To evaluate the localization of power within the designated latent channels, we conduct a swap test, as detailed in Algorithm \ref{algo:swap}. In this procedure, the first $k$ channels of a latent representation derived from an augmented signal (+6 dB) are replaced with the corresponding channels from the original signal. The resulting power ratio, $R_{\mathrm{dB}}$, of the decoded waveform serves as a measure of decoupling.

\begin{algorithm}[H]
\caption{Swap test for power localization in first $k$ latent channels}
\begin{algorithmic}[1]
\Require $x \in \mathbb{R}^N$; $\mathrm{Enc}: \mathbb{R}^N \to \mathbb{R}^{L \times T}$; $\mathrm{Dec}: \mathbb{R}^{L \times T} \to \mathbb{R}^N$

\State $Z \in \mathbb{R}^{L \times T} \gets \mathrm{Enc}(x)$
\State $Z' \in \mathbb{R}^{L \times T} \gets \mathrm{Enc}(2x)$ \Comment{+6 dB}
\State $Z'[:\!k, :] \gets Z[:\!k, :]$
\State $x' \in \mathbb{R}^N \gets \mathrm{Dec}(Z')$

\State $R_{\mathrm{dB}} \gets 10 \log_{10} \frac{\sum_{n=1}^N x'[n]^2}{\sum_{n=1}^N x[n]^2}$

\State \Return $R_{\mathrm{dB}}$
\end{algorithmic}
\label{algo:swap}
\end{algorithm}

An ideal result of $R_{\mathrm{dB}} \approx 0$ indicates that the power variation is effectively isolated within the first $k$ channels, as the swap successfully neutralizes the added gain. Conversely, $R_{\mathrm{dB}} \approx 6$ suggests that power information remains entangled across the remaining channels. In the latter case, the decoupling loss coefficient $\lambda_{\text{PoDAR}}$ must be increased to reduce the coupling between semantic and power information.

\section{Results}

\subsection{Codec reconstruction and swap test}
\label{ssec:codec}

Table~\ref{tab:combined_PoDAR_results} summarizes the reconstruction fidelity and swap-test results for the autoencoder representations described in Section \ref{ssec:autoencoder_baseline}.

\begin{table}[htbp]
  \caption{ViSQOL and swap-test gain per $\lambda_{\text{PoDAR}}$.}
  \label{tab:combined_PoDAR_results}
  \centering
  \begin{tabular}{@{}lccc@{}}
    \toprule
    VAE & $\lambda_{\text{PoDAR}}$ & ViSQOL & Swap Gain (dB) \\
    \midrule
    VAE1 (baseline) & 0.0  & $4.01 \pm 0.05$ & - \\
    VAE1 & 0.1  & $4.06 \pm 0.05$ & $+2.21 \pm 0.04$ \\
    \textbf{VAE1} & 0.5  & $3.70 \pm 0.04$ & $+0.99 \pm 0.02$ \\
    VAE1 & 0.75 & $3.65 \pm 0.05$ & $+0.97 \pm 0.02$ \\
    \midrule
    VAE2 (baseline) & 0.0  & $3.96 \pm 0.06$ & - \\
    \textbf{VAE2} & 0.1  & $3.96 \pm 0.06$ & $+0.83 \pm 0.02$ \\
    \bottomrule
  \end{tabular}
\end{table}

A 1 dB shift in signal amplitude is widely recognized as the just noticeable difference (JND) for human perception of audio loudness. Accordingly, the decoupling weight $\lambda_{\text{PoDAR}}$ is tuned for both representations to ensure that the measured swap gain remains below this perceptual threshold. Empirical evaluation demonstrates that the required $\lambda_{\text{PoDAR}}$ varies by architecture. For VAE2, a loss weight of 0.1 already achieves a swap gain below the 1 dB threshold while simultaneously preserving reconstruction quality. In contrast, this weight proves inadequate for the requirements of VAE1. While a coefficient of 0.5 or above reduced the swap gain below the 1 dB JND threshold, the reconstruction fidelity exhibits a clear monotonic decline as these weights are scaled.

The ViSQOL scores for VAE1 degrade significantly as $\lambda_{\text{PoDAR}}$ is scaled to 0.5. This reduction is anticipated, as the objective imposes explicit structural constraints on the latent space to enforce factor decoupling. However, as discussed in \secref{introduction}, superior reconstruction fidelity does not consistently translate to improved downstream generative performance \citep{xu2026makingreconstructionfidpredictive}. Consequently, the reduction in reconstruction quality is justified insofar as the enhanced modelability of the latent space translates to superior performance in the final generative model.

While the matched swap test in \tabref{combined_PoDAR_results} demonstrates that global gain variations are softly decoupled, achieving perfect disentanglement is not the objective of this framework. Phonetic content inherently dictates relative loudness (e.g., vowels are naturally louder than unvoiced fricatives), creating an intrinsic acoustic coupling. This is observable in non-matched swap tests, where swapping power channels between unrelated utterances reveals residual energy correlations tied to the original phonetic structure. Furthermore, as detailed in \appref{decoupling}, linear probing experiments confirm that some power information remains in the content channels. Ultimately, PoDAR achieves a soft decoupling that routes sufficient variance away from the semantic channels to make the distribution significantly easier for the downstream generative model to learn.

\subsection{Generator results}
\label{sec:generator_results}

We evaluate the generative performance of the PoDAR representation using the LibriSpeech and Seed-TTS benchmarks. \figref{all} illustrates the comparative trajectories for WER, speaker similarity, and UTMOS as defined in Section~\ref{ssec:metrics} for both baseline and PoDAR configurations across VAE1 and VAE2 as a function of training steps. The final performance for these models is summarized in Table~\ref{tab:main_results_table}.

\begin{table}[htbp]
\caption{Final WER, speaker similarity, and UTMOS performance across baseline and PoDAR configurations for VAE1 and VAE2 on the LibriSpeech and Seed-TTS benchmarks. \textbf{Bold} indicates the best within each dataset, VAE family, and metric.}
\centering
\small
\setlength{\tabcolsep}{6pt}
\renewcommand{\arraystretch}{1.05}
\begin{tabular}{lccc}
\toprule
Model & WER $\downarrow$ & speaker SIM $\uparrow$ & UTMOS $\uparrow$ \\
\midrule
\multicolumn{4}{l}{\textit{LibriSpeech-PC}} \\
\cmidrule(lr){1-4}
VAE1-Baseline      & $0.023 \pm 0.003$           & $0.592 \pm 0.005$           & $3.75 \pm 0.03$           \\
VAE1-PoDAR  & $0.023 \pm 0.003$  & $\mathbf{0.647 \pm 0.005}$  & $\mathbf{3.97 \pm 0.02}$  \\
VAE2-Baseline      & $0.024 \pm 0.003$  & $0.559 \pm 0.005$           & $4.10 \pm 0.02$           \\
VAE2-PoDAR  & $0.025 \pm 0.003$           & $\mathbf{0.595 \pm 0.005}$  & $\mathbf{4.17 \pm 0.01}$  \\
\midrule
\multicolumn{4}{l}{\textit{Seed-TTS (EN)}} \\
\cmidrule(lr){1-4}
VAE1-Baseline      & $0.017 \pm 0.003$           & $0.613 \pm 0.006$           & $3.69 \pm 0.02$           \\
VAE1-PoDAR  & $0.017 \pm 0.003$  & $\mathbf{0.660 \pm 0.005}$  & $\mathbf{3.72 \pm 0.02}$  \\
VAE2-Baseline      & $0.017 \pm 0.003$  & $0.594 \pm 0.006$           & $3.87 \pm 0.02$           \\
VAE2-PoDAR  & $0.019 \pm 0.003$           & $\mathbf{0.615 \pm 0.006}$  & $3.89 \pm 0.02$  \\
\midrule
\multicolumn{4}{l}{\textit{Seed-TTS (ZH)}} \\
\cmidrule(lr){1-4}
VAE1-Baseline      & $0.016 \pm 0.002$  & $0.713 \pm 0.003$           & $2.79 \pm 0.02$           \\
VAE1-PoDAR  & $0.017 \pm 0.002$           & $\mathbf{0.761 \pm 0.002}$  & $\mathbf{2.87 \pm 0.02}$  \\
VAE2-Baseline      & $0.020 \pm 0.002$           & $0.688 \pm 0.003$           & $2.98 \pm 0.02$  \\
VAE2-PoDAR  & $0.018 \pm 0.002$  & $\mathbf{0.701 \pm 0.003}$  & $2.96 \pm 0.02$           \\
\bottomrule
\end{tabular}
\label{tab:main_results_table}
\end{table}

The PoDAR representation facilitates accelerated convergence across all evaluated datasets, requiring approximately 2$\times$ fewer training steps to reach speaker similarity performance comparable to the baseline. Speaker similarity is notably higher for the generator trained on PoDAR, with a consistent improvement of 0.055 for VAE1 across both benchmarks and a slightly smaller gain for VAE2.

Regarding UTMOS, VAE1 demonstrates a 0.22 increase on LibriSpeech, while the improvement for VAE2 is more moderate, which we hypothesize stems from baseline regularization. As can be seen in \tabref{stable_audio_vaes}, VAE2 applies a KL penalty of $10^{-4}$, two orders of magnitude higher than VAE1  at $10^{-6}$, restricting the latent manifold for additional structure. Both architectures show only slight changes on the Seed-TTS dataset. These findings demonstrate that although the PoDAR codec exhibits lower reconstruction fidelity as described in Section~\ref{ssec:codec}, the resulting representation is more modelable and yields superior UTMOS scores compared to the baseline. Crucially, the contrasting results between the two architectures highlight that PoDAR can independently induce a highly modelable representation without relying on severe KL penalties (VAE1), while still providing consistent benefits atop already highly regularized spaces (VAE2).

Finally, the WER remains largely unchanged because the results for both configurations fall within the confidence intervals of the ground truth WER reported in \citet{chen2025f5tts}.

\section{Ablations}

All ablation experiments are conducted utilizing the VAE1 representation, with results reported on the LibriSpeech-PC dataset.

\subsection{Impact of partial CFG}

\begin{figure}[htbp]
    \centering    
    \includegraphics[width=0.9\textwidth]{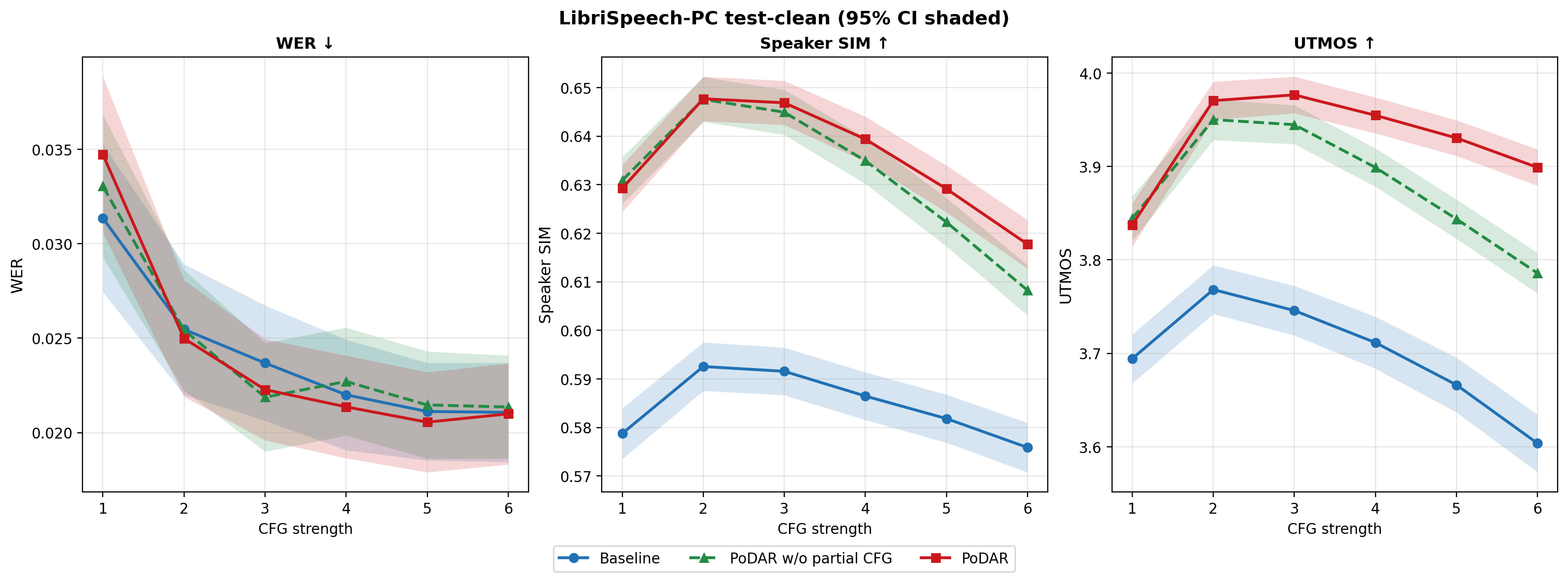}
    \caption{Impact of partial CFG on WER, speaker SIM and UTMOS performance across CFG scales.}
    \label{fig:partial}
\end{figure}

This ablation evaluates the impact of the partial CFG strategy within the PoDAR framework. We compare a configuration where CFG is applied across the full latent vector, including the power channels, against a selective approach where guidance is restricted to the power-invariant channels as described in Section~\ref{ssec:partial}. The performance metrics for both configurations are illustrated in \figref{partial}.

The PoDAR implementation without partial CFG already yields substantial performance gains over the baseline, in both speaker similarity and UTMOS metrics. While adding partial CFG achieves marginally higher peak values for both, the primary utility is its robustness at elevated guidance scales. Whereas the performance of the PoDAR implementation without partial CFG degrades rapidly as the CFG scale increases, the partial CFG strategy sustains high metric values across a broader range of guidance values.

\subsection{Impact of $\lambda\PoDAR$}

\begin{figure}[htbp]
    \centering
    \includegraphics[width=0.9\textwidth]{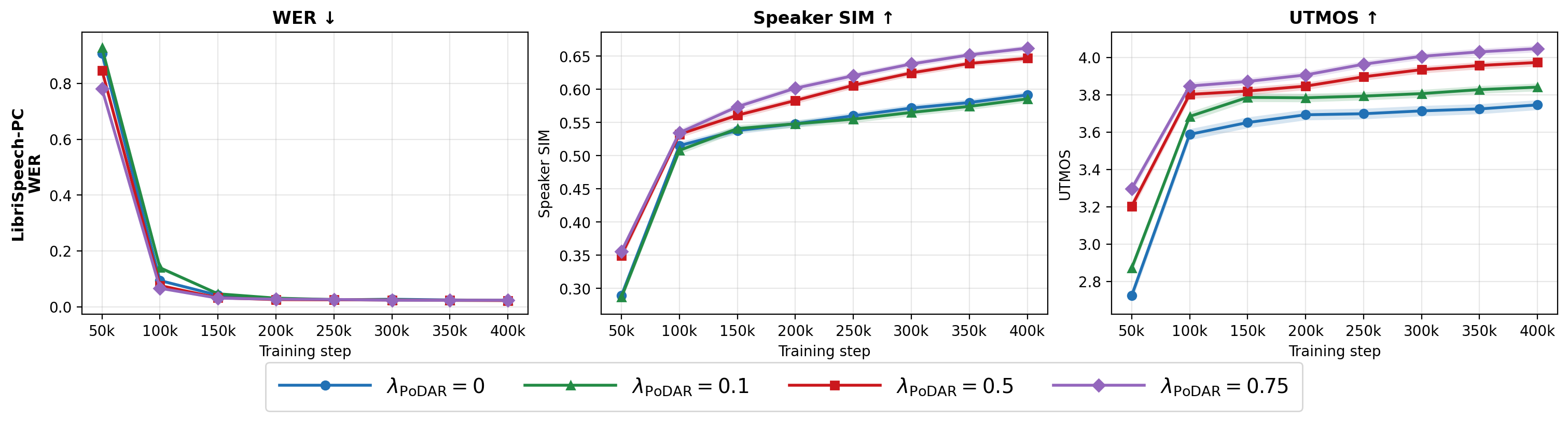}
    \caption{Impact of $\lambda\PoDAR$ on WER, speaker SIM, and UTMOS.}
    \label{fig:weights}
\end{figure}

This ablation evaluates generators trained on VAE1 representations with varying $\lambda_{\text{PoDAR}}$ coefficients detailed in Table~\ref{tab:combined_PoDAR_results}. We investigate how the strength of the decoupling objective during autoencoder training influences the performance of the downstream generator. The resulting comparisons across these configurations are illustrated in \figref{weights}.

Within the evaluated hyperparameter range, increasing $\lambda_{\text{PoDAR}}$ results in measurable improvements in speaker similarity and UTMOS metrics. A distinct divergence appears between the swap test results and the downstream generator performance. Although the configurations for $\lambda_{\text{PoDAR}} = 0.5$ and $\lambda_{\text{PoDAR}} = 0.75$ yielded comparable swap gain results in Table~\ref{tab:combined_PoDAR_results}, the latter achieves higher scores in speaker similarity and UTMOS. This finding suggests that the strength of the decoupling objective can influence generative performance beyond the attribution captured by the swap test.

\subsection{Impact of $k$}

A central design choice of PoDAR is the capacity allocated to the power subspace, determined by the number of power channels, $k$. Throughout the paper we utilize $k=1$, which, as shown in Section~\ref{ssec:codec}, empirically provided sufficient capacity to reduce the swap test error below the JND threshold of 1~dB. 

To verify this capacity requirement, we conducted an ablation using the VAE1 representation with $k=2$ power channels. The downstream model trained on the $k=2$ representation yielded UTMOS and speaker similarity metrics that were statistically indistinguishable from the $k=1$ configuration. These findings confirm that a single power channel provides sufficient representational capacity to capture the energy variance and allocating additional channels to the power subspace does not yield further generative improvements.

\section{Conclusion}
\label{sec:conclusion}
We introduced PoDAR (Power-Decoupled Audio Representation), a framework designed to enhance the modelability of continuous audio latent spaces by utilizing a randomized power augmentation and latent consistency objective to decouple signal power from invariant semantic content. Empirical evaluation on the LibriSpeech-PC benchmark demonstrates that PoDAR achieves about a $2\times$ acceleration in convergence to match baseline performance while concurrently providing gains of 0.055 in speaker similarity and 0.22 in UTMOS.
Moreover, this decoupling enables a partial CFG strategy that restricts semantic steering to the power-invariant subspace, effectively improving the robustness of the generator at elevated guidance scales.
\paragraph{Limitations.}
The primary limitations of this work involve the increased computational overhead during training of the autoencoders, since the consistency objective requires dual encoder passes for both the original sample $x$ and the augmented input $\mathcal{A}(x)$. Furthermore, the efficacy of the framework has been demonstrated exclusively within the speech domain and specifically targets power decoupling rather than broader acoustic factors or alternative modalities.
\paragraph{Future work.} While our downstream generative evaluation is strictly on TTS, the underlying PoDAR autoencoders are trained on a mixture of speech, music, and environmental audio. A natural extension of this research is to evaluate it on downstream text-to-audio (TTA) and text-to-music (TTM) tasks, where signals often exhibit even wider dynamic ranges. Future research may extend the augmentation-consistency principle to alternative modalities such as images and incorporate additional factors like voicing, pitch, saturation, or contrast by defining suitable augmentations and allocating dedicated latent subspaces for multi-factor decoupling. It remains an open question whether this augmentation-consistency framework can be applied to non-linear augmentations that introduce artifacts.

\subsection*{AI use statement}
In this work, we used generative AI tools to assist in implementing methods, using AI coding assistants to write and edit parts of our augmentation, loss and partial CFG code. We have not used generative AI tools for any other task in the required-disclosure category of the ICLR 2027 AI policy. Additionally, we used generative AI tools to draft parts of the paper, to edit the text for readability, to search for information, and to format references. We have reviewed and edited all AI-assisted work, including AI-generated code and AI-drafted or AI-edited text. We take responsibility for the final content of this work.

\subsection*{Ethics statement}
This work involves no human subjects, releases no data, models, or code, and raises no
privacy or data-security issues. All baselines, both the Stable Audio autoencoders and the F5-TTS generator, and all training and evaluation datasets are open source or publicly available and are used in accordance with their licenses, listed in Appendix~\ref{app:licenses} (Table~\ref{tab:licenses}). The main consideration is that by roughly halving the training needed to reach a given generation quality, PoDAR reduces the training compute (GPU-hours in Appendix~\ref{app:compute}) and with it the energy consumption and carbon footprint of training. However, the same efficiency lowers the barrier to misuse of speech generation, such as unauthorized voice cloning or synthetic-speech disinformation.
Finally, our evaluation covers only English and Mandarin, so we make no claims about
performance or fairness across other languages.

\subsection*{Reproducibility statement}
We do not release code, but the paper contains everything needed to reproduce our results.
The method consists of three self-contained modifications, each fully specified in
Section~\ref{sec:method}: the power augmentation,
the consistency objective in \eqref{lcons}, and the partial CFG rule in
\eqref{partialcfg}. Both autoencoder baselines and the generator are re-trained from
open-source implementations, \texttt{stable-audio-tools}
\citep{stabilityai_stableaudiotools} and F5-TTS \citep{chen2025f5tts}, following their
published recipes (Section~\ref{ssec:autoencoder_baseline}), and all training and
evaluation data are public datasets listed with their licenses in
Appendix~\ref{app:licenses}. Section~\ref{sec:experiments} gives the training mixtures,
data splits, test sets, and evaluation protocol, and Appendix~\ref{app:hparams} lists the training and inference hyperparameters of the baseline and PoDAR-modified autoencoders and of the latent generator, including the sampler configuration and random seed (Tables~\ref{tab:stable_audio_vaes} and~\ref{tab:hparams}).

% The authors would like to thank Descript for supporting and funding this research, including the provision of computational resources. We are especially grateful to Rachel Bloch Mellon for her leadership throughout this project.

\FloatBarrier
\newpage

\bibliographystyle{iclr2027_conference}
\bibliography{references}

\FloatBarrier
\newpage
%%%%%%%%%%%%%%%%%%%%%%%%%%%%%%%%%%%%%%%%%%%%%%%%%%%%%%%%%%%%
%%%%%%%%%%%%%%%%%%%%%%%%%%%%%%%%%%%%%%%%%%%%%%%%%%%%%%%%%%%%
\appendix

\section{Asset licenses}
\label{app:licenses}

Table~\ref{tab:licenses} lists every code repository, dataset, and pretrained
model used in this paper, together with its license and primary citation.
All assets are publicly available; no proprietary data or weights are used.

\begin{table}[htbp]
\caption{Licenses for all third-party assets used in this paper.}
\label{tab:licenses}
\centering
\small
\setlength{\tabcolsep}{4pt}
\begin{tabular}{@{}p{3.4cm}p{2.5cm}p{2.1cm}p{5.0cm}@{}}
\toprule
Asset & Type & License & Reference \\
\midrule
\multicolumn{4}{l}{\textit{Code / pretrained models}} \\
\cmidrule(lr){1-4}
\texttt{stable-audio-tools}            & VAE training code      & MIT                & \citealp{stabilityai_stableaudiotools} \\
F5-TTS                                  & Generator code        & MIT                & \citealp{chen2025f5tts} \\
faster-whisper (large-v3)              & ASR (en WER)          & MIT                & \citealp{radford2022whisper} \\
FunASR \texttt{paraformer-zh}          & ASR (zh WER)          & MIT                & \citealp{gao2023funasrfundamentalendtoendspeech} \\
WavLM-Large \mbox{+ ECAPA-TDNN}               & SIM-o backbone        & MIT/Apache 2.0                & \citealp{chen2022wavlm,desplanques2020ecapatdnn} \\
SpeechMOS \texttt{utmos22\_strong}     & UTMOS predictor       & MIT                & \citealp{saeki2022utmos} \\
\midrule
\multicolumn{4}{l}{\textit{Autoencoder training datasets (DAC mixture)}} \\
\cmidrule(lr){1-4}
DAPS                                    & Speech                & CC BY-NC 4.0       & \citealp{mysore2015daps} \\
DNS Challenge 4 (clean speech)         & Speech                & CC BY 4.0          & \citealp{dubey2022dns} \\
Common Voice                            & Speech                & CC0 1.0            & \citealp{ardila2019commonvoice} \\
VCTK                                    & Speech                & ODC-By 1.0         & \citealp{yamagishi2019vctk} \\
MUSDB18                                 & Music                 & CC BY-NC-SA 4.0    & \citealp{rafii2017musdb18} \\
MTG-Jamendo                             & Music                 & CC BY-NC-SA 4.0    & \citealp{bogdanov2019mtgjamendo} \\
AudioSet (bal.\ + unbal.\ train)       & Environmental         & CC BY 4.0          & \citealp{gemmeke2017audioset} \\
\midrule
\multicolumn{4}{l}{\textit{Generator training dataset}} \\
\cmidrule(lr){1-4}
Emilia ZH-EN                            & Speech (bilingual)    & CC BY-NC 4.0       & \citealp{he2024emilia} \\
\midrule
\multicolumn{4}{l}{\textit{Evaluation datasets}} \\
\cmidrule(lr){1-4}
LibriSpeech (test-clean)               & Speech (en)           & CC BY 4.0          & \citealp{panayotov2015librispeech} \\
LibriSpeech-PC (cross-sentence)   & Speech (en)           & CC BY 4.0          & \citealp{meister2023librispeechpc} \\
Seed-TTS test-en                        & Speech (en)           & CC BY-NC 4.0  & \citealp{anastassiou2024seedttsfamilyhighqualityversatile} \\
Seed-TTS test-zh / \mbox{DiDiSpeech}          & Speech (zh)           & CC BY-NC 4.0  & \citealp{anastassiou2024seedttsfamilyhighqualityversatile,guo2021didispeechlargescalemandarin} \\
\bottomrule
\end{tabular}
\end{table}

\section{Compute resources}
\label{app:compute}

All training experiments were conducted using 4 NVIDIA H200 GPUs and the same hardware was used for inference and metric evaluation. Table \ref{tab:compute} summarizes the per run and aggregate compute footprint for the training results reported in this paper, totaling approximately 4,608 GPU hours.

\begin{table}[htbp]
\caption{Training-compute footprint reported in this paper.
\emph{Codecs} covers the 4 autoencoders used for the main results
(VAE1 / VAE2 $\times$ baseline / PoDAR) plus the 2 PoDAR-weight ablations
on VAE1 ($\lambda_{\text{PoDAR}} \in \{0.1, 0.75\}$).
\emph{Generators} covers one F5-TTS v1 Base latent generator per codec
($400{,}000$ optimizer updates each).}
\label{tab:compute}
\centering
\small
\setlength{\tabcolsep}{6pt}
\begin{tabular}{@{}lcccc@{}}
\toprule
Phase & Per-run wall-clock & Runs & GPUs/run & Subtotal (GPU-h) \\
\midrule
Codec training (1 epoch on DAC mixture)
  & $\sim 1$\,day  & 6 & 4~$\times$~H200 & $576$ \\
Generator training ($400{,}000$ updates)
  & $\sim 7$\,days & 6 & 4~$\times$~H200 & $4{,}032$ \\
\midrule
\textbf{Training total}
  & ---            & --- & ---             & $\mathbf{4{,}608}$ \\
\bottomrule
\end{tabular}
\end{table}

We estimate that the total computational budget for the full research cycle including inference runs and unsuccessful iterations was below 10,000 GPU hours.

\section{Full training hyperparameters}
\label{app:hparams}

Table \ref{tab:stable_audio_vaes} summarizes the structural differences and optimization strategies for the baseline Stable Audio autoencoders utilized throughout this work.

\begin{table}[htbp]
  \caption{Stable Audio autoencoder baselines from \texttt{stable-audio-tools}.}
  \label{tab:stable_audio_vaes}
  \centering
  \small
  \begin{tabular}{@{}p{3.3cm}p{4.0cm}p{4.0cm}@{}}
    \toprule
    Aspect & Stable Audio 1.0 VAE & Stable Audio 2.0 VAE \\
    \midrule
    Backbone & \textbf{DAC} & \textbf{Oobleck} \\
    Encoder strides & \texttt{[4, 4, 8, 8]} & \texttt{[2, 4, 4, 8, 8]} \\
    Output nonlinearity & \textbf{Tanh} & \textbf{None} \\
    Discriminator filters & \texttt{32} & \texttt{64} \\
    KL weight & \texttt{1e-6} & \texttt{1e-4} \\
    AE optimizer & LR \texttt{1e-4}, no weight decay & LR \texttt{1.5e-4}, weight decay \texttt{1e-3} \\
    LR schedule & ExponentialLR (\texttt{gamma=0.999996}) & InverseLR (\texttt{inv\_gamma=200000}, \texttt{power=0.5}, \texttt{warmup=0.999}) \\
    Discriminator & LR \texttt{1e-4} & LR \texttt{3e-4} \\
    \bottomrule
  \end{tabular}
\end{table}

Additionally, we provide the training configurations necessary for the full reproduction of our experimental results in this section. Table \ref{tab:hparams} details the specific hyperparameters utilized for both the latent generator and the PoDAR modified autoencoder architectures including the specialized power augmentation settings.

\begin{table}[htbp]
\caption{Generator and autoencoder training hyperparameters. Generator
configuration mirrors the published F5-TTS v1 Base recipe with input/output
projections resized to the autoencoder latent dimension.}
\label{tab:hparams}
\centering
\small
\begin{tabular}{@{}lll@{}}
\toprule
Component & Hyperparameter & Value \\
\midrule
\multicolumn{3}{l}{\textit{F5-TTS v1 Base latent generator}} \\
\cmidrule(lr){1-3}
Backbone     & Architecture                       & DiT \\
             & Hidden dim / depth / heads         & 1024 / 22 / 16 \\
             & FF mult / text dim                 & 2 / 512 \\
             & ConvNeXt-V2 text layers            & 4 \\             
Conditioning & $p_{\text{audio drop}}$            & 0.3 \\
             & $p_{\text{uncond}}$                & 0.2 \\
Optim        & Optimizer                          & fused AdamW \\
             & Learning rate (peak)               & $7.5\times 10^{-5}$ \\
             & Weight decay                       & 0 \\
             & Grad clip ($\ell_2$)               & 1.0 \\
             & Warmup updates                     & 20{,}000 \\
             & Schedule                           & linear warmup, linear decay \\
             & Mixed precision                    & fp16 \\
             & EMA                                & 0.9999 \\
Sampling     & ODE solver                         & Euler \\
             & NFE                                & 32 \\
             & Sway sampling                      & $-1$ \\
             & Seed                               & 0 \\
             & CFG scale                          & 3.0 (VAE1), 2.0 (VAE2) \\
\midrule
\multicolumn{3}{l}{\textit{Stable Audio 1.0 / 2.0 VAE (PoDAR-modified)}} \\
\cmidrule(lr){1-3}
Data         & Segment length                     & $65{,}536$ samples (1.49\,s) \\
             & Sample rate                        & 44{,}100\,Hz \\
             & Mixture                            & DAC training mix (Section \ref{ssec:autoencoder_baseline}) \\
PoDAR        & Power channels $k$                 & 1 \\
             & $\lambda_{\text{PoDAR}}$ (VAE1)   & 0.5 (main); 0.1, 0.75 (ablation) \\
             & $\lambda_{\text{PoDAR}}$ (VAE2)   & 0.1 \\
             & Augmentation                       & uniform gain in $[-6,+6]$\,dB \\
\bottomrule
\end{tabular}
\end{table}

\section{Latent variance analysis}
\label{app:variance}

To quantify the impact of the $L_2$ consistency objective ($\mathcal{L}\PoDAR$) on the latent space, we measured both the scale of the latent manifold (the standard deviation of the predicted means, $\mu$) and the posterior uncertainty (the mean of the predicted standard deviations, $\sigma$) on the DAPS clean test set. The results are summarized in \tabref{latent_variance}.

\begin{table}[htbp]
\centering
\small
\caption{Impact of the $\mathcal{L}\PoDAR$ objective on latent manifold statistics measured on the DAPS clean test set.}
\label{tab:latent_variance}
\begin{tabular}{@{}lcc@{}}
\toprule
Codec & Posterior std (mean of $\sigma$) & Signal scale (std of $\mu$) \\
\midrule
VAE1-Baseline & $0.006$ & $1.482$ \\
VAE1-PoDAR    & $0.002$ & $0.395$ \\
VAE2-Baseline & $0.120$ & $0.993$ \\
VAE2-PoDAR    & $0.124$ & $0.863$ \\
\bottomrule
\end{tabular}
\end{table}

The $\mathcal{L}\PoDAR$ objective inherently regularizes the overall latent space, but the magnitude of this effect is highly dependent on the baseline KL penalty. For VAE1, which employs a low KL penalty ($10^{-6}$), the initially unconstrained baseline manifold shrinks significantly, with the signal scale dropping from 1.482 to 0.395. Conversely, for VAE2, which utilizes a much higher KL penalty ($10^{-4}$), the manifold is already tightly constrained, so the additional $L_2$ loss exerts a much smaller relative impact on the signal scale, from 0.993 to 0.863.

Crucially, these statistics justify our choice in \secref{method} to apply $\mathcal{L}\PoDAR$ exclusively to the means ($\mu$) rather than the standard deviations ($\sigma$). For VAE1, the posterior $\sigma$ is 0.006, practically zero, rendering an $L_2$ penalty applied to it meaningless. For VAE2, the high KL penalty independently anchors the posterior $\sigma$, keeping it stable at approximately 0.12 even with the addition of PoDAR. Therefore, applying the consistency penalty to the variance parameter is unnecessary across both operating regimes.

\section{Strictness of representation decoupling}
\label{app:decoupling}

As discussed in Section~\ref{ssec:codec}, our goal is soft decoupling rather than perfect disentanglement, given that acoustic semantic content and energy are intrinsically coupled. To evaluate the strictness of the separation between the power and semantic subspaces, we trained a linear probe to predict the power channel using solely the remaining semantic channels as input. The performance metrics for this evaluation are summarized in \tabref{linear_probe}.

\begin{table}[htbp]
\centering
\small
\caption{Linear probe performance predicting the power channel from the semantic channels.}
\label{tab:linear_probe}
\begin{tabular}{@{}lcc@{}}
\toprule
Representation & $R^2$ & Pearson $r$ \\
\midrule
VAE1-PoDAR & $0.67$ & $0.83$ \\
VAE2-PoDAR & $0.66$ & $0.81$ \\
\bottomrule
\end{tabular}
\end{table}

These results show that, while PoDAR successfully routes enough energy variance into the power subspace to improve latent modelability, the semantic channels retain phonetically-linked power information.

\section{Latent-space smoothness}
\label{app:latent-smoothness}

As noted in Section~\ref{ssec:metrics}, we also probe each codec by perturbing its latents and measuring how much the decoded audio degrades. The intuition behind this is that a latent space is more modelable if it is smooth, i.e. if nearby latents decode to perceptually similar audio. The more tolerant it is to local perturbations, the easier the task of the downstream generative model. These geometric probes capture large differences in latent geometry, but they are not sensitive enough to serve as a standalone modelability metric, so we report them here as a diagnostic.

\paragraph{Data and segmentation.}
We evaluate on the clean subset of the DAPS corpus. Each recording is converted to mono, resampled to the codec sample rate ($44.1$~kHz), and power-normalised to $-16$~dB. From every file we take four non-overlapping $10$~s segments starting $5$~s into the recording, giving $40$ segments in total; each segment is zero-padded to a multiple of the encoder downsampling ratio. All four codecs see the same segments, and all randomness is seeded per segment.

\paragraph{Covariance-shaped perturbations.}
Let $\vec{Z} \in \mathbb{R}^{C \times T}$ be the latent of one segment, with $C$ channels and $T$ frames, and let $\Sigma = \mathrm{Cov}(\vec{Z}) \in \mathbb{R}^{C \times C}$ be its empirical channel covariance across time, with eigendecomposition $\Sigma = V \diag(\lambda) V^{\!\top}$. We define the square-root factor $A = V \diag(\lambda^{1/2})$ and perturb the latent as
\begin{equation}
  \tilde{\vec{Z}}(\sigma) = \vec{Z} + \sigma \, A \vec{E},
  \qquad E_{c,t} \sim \mathcal{N}(0, 1) \ \text{i.i.d.}
  \label{eq:perturb}
\end{equation}
so that the injected noise has covariance $\sigma^2 \Sigma$ per frame. Shaping the noise by the codec's own latent covariance makes a perturbation of magnitude $\sigma$ correspond to the same \emph{relative} displacement along every principal direction of the latent distribution, independent of the absolute scale or anisotropy of a particular codec.

\paragraph{Distortion metric.}
Degradation is measured by mel-cepstral distortion (MCD). Mel spectrograms use $96$ mel bands with hop length equal to the encoder downsampling ratio and window length twice the hop ($50\%$ overlap), so that one mel frame still corresponds to exactly one latent frame. We use 13 cepstral coefficients $c_1,\dots,c_{13}$, excluding $c_0$ so that a constant loudness offset does not penalise every frame, and flooring the mel magnitudes at $10^{-5}$ so that silent frames do not dominate:
\begin{equation}
  \mathrm{MCD} = \frac{10}{\ln 10} \cdot
  \frac{1}{T} \sum_{t=1}^{T}
  \sqrt{2 \sum_{k=1}^{13} \bigl(c^{(1)}_{k,t} - c^{(2)}_{k,t}\bigr)^2 } .
\end{equation}

\begin{figure}[htbp]
    \centering
    \includegraphics[width=0.85\textwidth]{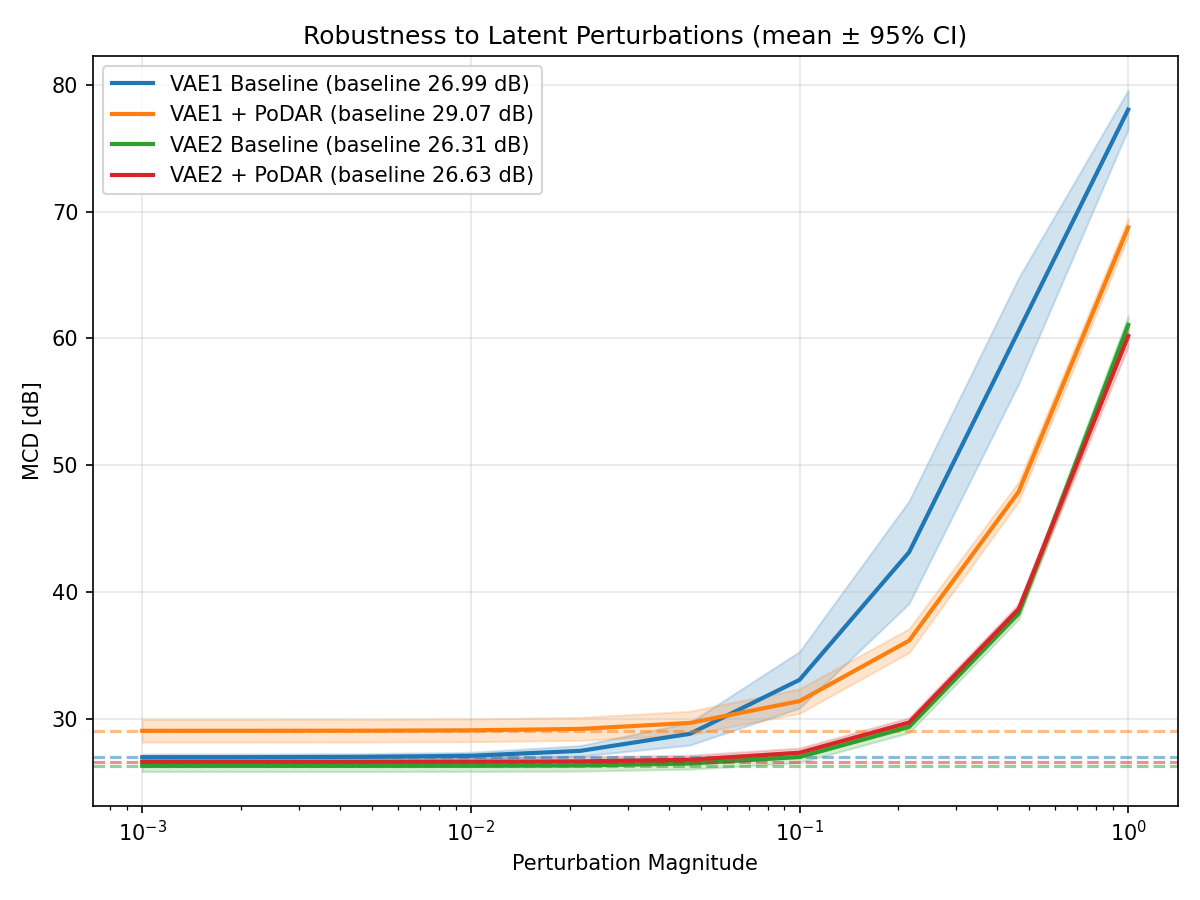}
    \caption{Robustness to covariance-shaped latent perturbations. MCD is measured between the original waveform and the decode of the perturbed latent, as a function of perturbation magnitude $\sigma$. Dashed lines are unperturbed reconstruction MCD. Mean $\pm$ 95\% CI over 40 DAPS segments.}
    \label{fig:smoothness}
\end{figure}

\paragraph{Smoothness (robustness) curve.}
For each segment we sweep $\sigma$ over $10$ logarithmically spaced values in $[10^{-3}, 10^{0}]$, decode $\tilde{\vec{Z}}(\sigma)$ from Eq.~\eqref{perturb}, and report the MCD between the decoded audio and the original waveform. The dashed horizontal lines in \figref{smoothness} give the unperturbed reconstruction MCD of each codec, i.e.\ the floor that the curve approaches as $\sigma \to 0$. These floors are relatively high because codec reconstruction is imperfect, particularly on some DAPS segments, so even $\sigma=0$ already incurs substantial MCD relative to the original waveform. A curve that stays flat over a wider range of $\sigma$ before rising indicates a smoother latent space: nearby latents decode to perceptually similar audio, so a downstream generator operating in that space is less sensitive to prediction error. Shaded bands are $95\%$ confidence intervals of the mean over the $40$ segments. The ranking in \figref{smoothness} is VAE2-PoDAR slightly more robust than VAE2-Baseline, which is more robust than VAE1-PoDAR, which is in turn substantially more robust than VAE1-Baseline. This ordering matches the UTMOS trend (and the gaps between the performance of the different codecs) in \tabref{main_results_table} for LibriSpeech-PC but it is not very conclusive for the other datasets.

\end{document}